\documentclass[12pt,english]{article}
\usepackage{babel}
\usepackage{graphicx}
\usepackage{amsmath}
\usepackage{bm}
\usepackage{latexsym}

\begin{document}

\title{Aproximative solutions to the neutrino oscillation problem in matter}
\author{Ver\'{\i}ssimo M. de Aquino and J. S. S. de Oliveira \\
Universidade Estadual de Londrina}
\maketitle

\begin{abstract}
We present approximative solutions to the neutrino evolution equation
calculated by different methods. In a two neutrino framework, using the
physical parameters which gives the main effects to neutrino oscillations
from $\nu _{e}$ to another flavors for $L\simeq 3000Km$ and $E\simeq 1GeV$ ,
the results for the transition probability calculated by using series
solutions, by to take the neutrino evolution operator as a product of
ordered partial operators \ and by numerical methods, for a linearly and
sinusoidally varying matter density are compared. The extension to an
arbitrary density profile is discussed and the evolution operator as a
product of partial operators in the three neutrino case is obtained.
\end{abstract}

\section{Introduction}

The conversion of one type of neutrino into another, while propagating in
vacuum or in matter, can be the solution to the Solar and atmospheric
neutrino problems and to explain the results about neutrino flux from
experiments with reactors and accelerators. In order for such vacuum or
matter oscillations to occur, it is necessary that the neutrino have a
nonzero mass and that the neutrino masses be not all degenerate. These
possibilities have been explored in several models like that ones where a
term of mass for neutrinos is constructed by the introduction of new Higgs
fields and the mass is generated by spontaneous breakdown of symmetry,
models with fermion sector extended and more complex models. For a brief
review about these models, see [1]. In this context flavour eigenstate of
neutrinos can be a mixing of mass eigenstates and oscillations can take
place. Independent of specific models, for a neutrino initially created as a
flavour eigenstate, the amplitude of probability for conversion to another
flavor and the survival amplitude of probability, after to travel a distance
L, in vacuum or in matter [2], are given by the solutions of the evolution
equation. If CP conservation is assumed the evolution equation depends on
five parameters, three mixing angles and two differences of squared masses.
There is exact solution to the evolution equation for neutrinos propagating
in vacuum or in a constant matter density in a two neutrino framework [3].
The three neutrino case in a constant matter density has been analyzed in
the references [4],[5],[6]. The exact expression to the evolution operator
for a three neutrino system in a constant matter density in terms of the
mass squared differences and the vacuum mixing angles has been firstly
derived in the reference [5], the effective mixing angles in matter have
been also derived. In the reference [6] an alternative expression has been
derived. For varying matter density solutions can be found in the adiabatic
approximation [7 ] and in the nonadiabatic case near the resonances [8].
These solutions can give a estimation about the flux of neutrinos in the
cases which the neutrino source is big and there is a continuum spectrum of
energy, the integral on the energy or on the point of neutrino production
cancels contributions from the interference terms. For situations where the
dimension of the source is shorter than the oscillation lengths, for
determined neutrino energy, numerical solutions for the evolution equation
can give informations about mixing angles and differences of squared masses.
This is the case for neutrinos from reactors and accelerators and
atmospheric neutrinos travelling across the earth at energy $E\approx 1GeV$
. The evolution equation for a neutrino system propagating at variable
matter density has recently been studied . In the reference [9], in a two
neutrino framework, the strategy is to expand the neutrino oscillation
probability and the effective potential in series. \ In the reference [10]
the strategy of Elisio Lisi and Daniele Montanino is to take the evolution
operator given by the ordered product of partial evolution operators from
the shells along of the neutrino trajectory. In many of the three flavour
descriptions of neutrino oscillations the so called 'one mass scale
dominance' $\delta m^{2}=\Delta m_{21}^{2}<<\Delta m_{31}^{2}\simeq \Delta
m_{32}^{2}$ has been adopted. In these cases, for particular situations, the
problem reduce to the two neutrinos one. This is the case of the reference
[11] where the simple approach $U_{13}\approx 1$ has been used and in the
ref. [12] where the influence of the mixing angle $\theta _{13}$ has been
tacked into account for Sub-GeV atmospheric neutrinos and the accuracy of
the approximation is determined by the relation $tg2\widetilde{\theta }%
_{13}\simeq 2\widetilde{\theta }_{13}$. In these cases evolution equations
have been derived and the solutions in a varying matter case needs to be
found with the help of some approximative method. In ref. [13], in a two
neutrino scenario, approximate solution to the evolution equation has been
presented for high energetic neutrinos, $E>10$ $GeV$ for neutrinos
propagating at the matter of the Earth. In ref. [14] approximate solution to
the neutrino equation in a two neutrino framework with corrections due to
the three neutrino mixing has been presented for low energetic neutrinos ,
the validity of the solution depends on the parameter $\epsilon =\frac{2EV(x)%
}{\Delta m^{2}}$. The main problem of to take the evolution operator as a
ordered product of partial operators in the three neutrino case is that the
expression to the evolution operator in this case is not simple. In
reference [15] this strategy has been adopted but the approach $\Delta
m_{21}^{2}=0$ reduce in some sense the complications to that ones from the
two neutrino case. In reference [16] the general case has been analyzed
using the results from [6]. In this work, in section II, in order to obtain
numerical solutions by methods like Runge Kuta, the evolution equation of
the neutrino system at variable electron density is written as a system of
two disconnected second order linear ordinary differential equations. The
conversion probability is computed for neutrinos travelling at linear matter
density. In order to compare our results with another ones the parameters of
the system are those used in reference [9]. \ In section III the evolution
operator of the neutrino system at variable electron density is computed as
the product of partial operators. The conversion probability is computed for
the case analyzed in section II and we apply the methods to compute the
conversion probability for neutrinos travelling in a sinusoidally varying
matter density. In section IV we apply the strategy of to take the evolution
operator as a product of ordered partial operators to the three neutrino
case using the results from reference [5]. The conversion probabilities are
computed for the constant and the linearly varying matter density cases
Finally, in section V, we summarize our results and give our conclusions.

\section{Neutrinos in matter at variable density}

\qquad In a two neutrino framework the evolution equation for a neutrino
system propagating at matter is given by%
\begin{equation}
i\frac{d}{dx}\nu _{f}(x)=H(x)\nu _{f}(x)
\end{equation}%
where
\begin{equation}
\nu _{f}(x)=\left(
\begin{tabular}{l}
$\nu _{e}(x)$ \\
$\nu _{\mu }(x)$%
\end{tabular}%
\right) \text{,}
\end{equation}%
$H(x)=A_{3}\sigma _{3}+A_{1}\sigma _{1}$ and we have defined $A_{1}=\frac{1}{%
4E}\Delta \sin 2\theta $ and $A_{3}=\frac{1}{4E}(A(x)-\Delta \cos 2\theta )$
, $\sigma _{3}$ and $\sigma _{1}$ are Pauli matrices and in these
expressions, $A(x)=2\sqrt{2}G_{F}N_{e}(x)E$ , $\Delta =m_{2}^{2}-m_{1}^{2}$
denote the mass squared difference between the two mass eigenstates in
vacuum, $\theta $ is the mixing angle and $N_{e}(x)$ and $E$ are the
electron density and neutrino energy respectively. Introducing the new
fields $\Phi (x)$ by the relation
\begin{equation}
\nu _{f}(x)=e^{-i\int\limits^{x}d\lambda A(\lambda )\sigma _{3}}\Phi (x)%
\text{,}
\end{equation}%
using the auxiliary matrices $\sigma _{\pm }=\frac{\sigma _{1}\pm i\sigma
_{2}}{2}$, the definition $\beta (x)=\int\limits^{x}d\lambda A(\lambda )$,
and with the help of the Backer-Hausdorff relation for exponential operators%
\begin{equation}
e^{-F}Ge^{F}=\sum\limits_{n=0}^{\infty }\frac{G_{n}}{n!}%
;G_{0}=G;G_{n}=[G_{n-1},F]
\end{equation}
the evolution equation, in terms of the field $\Phi (x)=\left(
\begin{tabular}{l}
$\phi _{1}(x)$ \\
$\phi _{2}(x)$%
\end{tabular}%
\right) $, reads%
\begin{equation}
i\frac{d}{dx}\Phi (x)=A_{1}[e^{i2\beta (x)}\sigma _{+}+e^{-i2\beta
(x)}\sigma _{-}]\Phi (x)
\end{equation}%
or%
\begin{equation}
i\frac{d}{dx}\phi _{1}(x)=A_{1}e^{i2\beta (x)}\phi _{2}(x)
\end{equation}%
\begin{equation*}
i\frac{d}{dx}\phi _{2}(x)=A_{1}e^{-i2\beta (x)}\phi _{1}(x)
\end{equation*}%
solving this equations for the fields $\phi _{i}(x)$ one obtain the
following linear system of second order differential equations [17]
\begin{equation}
\frac{d^{2}\phi _{1}(x)}{dx^{2}}-2iA_{3}(x)\frac{d\phi _{1}(x)}{dx}%
+A_{1}^{2}(x)\phi _{1}(x)=0
\end{equation}%
\begin{equation*}
\frac{d^{2}\phi _{2}(x)}{dx^{2}}+2iA_{3}(x)\frac{d\phi _{2}(x)}{dx}%
+A_{1}^{2}(x)\phi _{2}(x)=0.
\end{equation*}%
For neutrinos produced as $\nu _{e}$ for example, the initial conditions to
the fields\ $\phi _{i}(x)$ becomes
\begin{equation}
\phi _{1}(0)=1;\phi _{1}^{\prime }(0)=0;\phi _{2}(0)=0;\phi _{2}^{\prime
}(0)=-iA_{1}
\end{equation}%
and the equations (7) with the initial conditions (8) can be solved by using
some numerical method for an arbitrary matter density. The survival and
conversion amplitudes of probability are given by the expressions
\begin{eqnarray}
\nu _{e}(x) &=&e^{-i\beta (x)}\phi _{1}(x) \\
\nu _{\mu }(x) &=&e^{i\beta (x)}\phi _{2}(x)  \notag
\end{eqnarray}%
We apply this results for \ a linearly varying matter density $%
N_{e}(x)=N_{core}(1+\frac{x}{L})/4$ where $N_{core}$ is the electron density
in the earths' core and $L=3000$ $km$. In the numerical calculations, we
have used the mixing angle $\theta =13^{0}$ and the squared mass difference $%
\Delta m^{2}=2\times 10^{-3}eV^{2}$. These choices are made in order to our
results can be compared with the results from [9]. The solutions has been
found numerically. In Fig. 1 the transition probability as a function of
distance is presented.

\section{The evolution operator}

The solution of (1) in a general way is $\nu _{f}(x)=U(x,x_{0})\nu
_{f}(x_{0})$ where $U(x,x_{0})$ is the evolution operator of the neutrino
system given explicitly by%
\begin{equation}
U(x,x_{0})=Exp\left\{ -i\int\limits_{x_{0}}^{x}H(\lambda )d\lambda \right\}
\end{equation}%
where $H$\ is the Hamiltonian of the system and the symbol Exp concern to
expansional that represent a expansion in orderly multiple integrals%
\begin{equation}
Exp\left\{ -i\int\limits_{x_{0}}^{x}H(\lambda )d\lambda \right\}
=1-i-i\int\limits_{x_{0}}^{x}H(\lambda )d\lambda
+(i)^{2}\int\limits_{x_{0}}^{x}H(\lambda )d\lambda
-i\int\limits_{x_{0}}^{\lambda }H(\lambda ^{\prime })d\lambda ^{\prime
}+....
\end{equation}%
and that for $\left[ H(x),H(x^{\prime })\right] =0$ is reduced to usual
exponential. This is the case of neutrinos travelling the vacuum and the
matter of constant density. The strategy of Elisio Lisi and Daniele
Montanino [10] is to divide the neutrino trajectory in infinitesimal
elements j in which the electronic density is given by $N_{j}(x)=\overline{%
N_{j}}(x)+\delta N_{j}(x)$%
\begin{equation}
\overline{N_{j}}(x)=\frac{1}{x_{j}-x_{j-1}}\int%
\limits_{x_{j-1}}^{x_{j}}N_{e}(x)dx
\end{equation}%
and the neutrino propagation is described by the Hamiltonian $H(x)=\overline{%
H_{j}}+\delta H_{j}(x)$. In this way, the evolution operator is taken as an
product of ordered partial operators%
\begin{equation}
U(x,x_{0})=\prod\limits_{j}U_{j}(x_{j},x_{j-1})
\end{equation}%
Note that for a linearly varying matter density the quantities $\overline{%
N_{j}}(x)$ and $\overline{H_{j}}(x)$ correspond to $N_{e}(\overline{x})$ and
$H(\overline{x})$ where $\overline{x}=\frac{x_{j}+x_{j-1}}{2}$.

The Hamiltonian\ $H_{j}(x)$, except for the proportional term to the
identity that just introduces in the evolution operator a constant phase
factor is given by%
\begin{equation}
H_{j}(x)=\frac{1}{2}\left\{
\begin{tabular}{ll}
$\sqrt{2}G_{F}\overline{N_{j}}-k\cos 2\theta $ & $k\sin 2\theta $ \\
$k\sin 2\theta $ & $-\sqrt{2}G_{F}\overline{N_{j}}+k\cos 2\theta $%
\end{tabular}%
\right\} +\delta H_{j}(x)
\end{equation}%
where $k=\frac{\Delta m^{2}}{2E}$ and $\delta H_{j}(x)=\frac{\sqrt{2}%
G_{F}\delta N_{j}(x)}{2}\sigma _{3}$.

The operator in the jth shell is%
\begin{equation}
U_{j}(x_{j},x_{j-1})=Exp\left\{ -i\int\limits_{x_{j-1}}^{x_{j}}[\overline{%
H_{j}}+\delta H_{j}(x)]dx\right\} .
\end{equation}%
With the help of the relation between exponential operators [18]%
\begin{equation}
Exp\left\{ -i\int\limits_{\sigma }^{\tau }[F(\lambda )+G(\lambda )]d\lambda
\right\} =Exp\left\{ -i\int\limits_{\sigma }^{\tau }F(\lambda )d\lambda
\right\} \ast
\end{equation}%
\begin{equation*}
Exp\left\{ -i\int\limits_{\sigma }^{\tau }E_{1}(\lambda )G(\lambda
)E_{2}(\lambda )d\lambda \right\}
\end{equation*}%
where%
\begin{equation*}
E_{1}(\lambda )=Exp\left\{ -i\int\limits_{\lambda }^{\sigma }F(u)du\right\}
\end{equation*}%
\begin{equation*}
E_{2}(\lambda )=Exp\left\{ -i\int\limits_{\sigma }^{\lambda }F(u)du\right\}
\end{equation*}%
and by note that $\overline{H_{j}}$ is constant, the evolution operator $%
U_{j}(x_{j},x_{j-1})$ becomes%
\begin{equation}
U_{j}(x_{j},x_{j-1})=e^{-i\overline{H_{j}}(x_{j}-x_{j-1})}\ast Exp\left\{
-i\int\limits_{x_{j-1}}^{x_{j}}e^{-i\overline{H_{j}}(x_{j-1}-x)}\delta
H_{j}(x)e^{-i\overline{H_{j}}(x-x_{j-1})}dx\right\}
\end{equation}%
and this expression permit us to take the evolution operator in orders of
perturbation. At the first order $U_{j}(x_{j},x_{j-1})$ results%
\begin{equation}
U_{j}(x_{j},x_{j-1})=e^{-i\overline{H_{j}}(x_{j}-x_{j-1})}-i\int%
\limits_{x_{j-1}}^{x_{j}}e^{-i\overline{H_{j}}(x_{j}-x)}\delta H_{j}(x)e^{-i%
\overline{H_{j}}(x-x_{j-1})}dx
\end{equation}%
where the first term corresponds to \ the operator for a neutrino travelling
at a constant matter density. The expression to the zero order term can be
obtained easily, but, in order to prepare the method to the three neutrino
case, when this expression is more complicated, we summarize the steps for
to do this. The way is to calculate the eigenvalues $\lambda _{1,2}$ to the
Hamiltonian $\overline{H_{j}}$ , $\lambda _{1,2}=\frac{\overline{\omega }}{4E%
}$ where $\varpi =\sqrt{\Delta ^{2}+A^{2}-2A\Delta \cos 2\theta }$ and to
obtain the expression to the evolution operator in the particle basis. The
result is
\begin{equation}
\overline{U}_{j}^{P}(x_{j},x_{j-1})=\left\{
\begin{tabular}{ll}
$e^{i\varpi /4E}$ & $0$ \\
$0$ & $e^{-i\varpi /4E}$%
\end{tabular}%
\right\}
\end{equation}%
Then, the evolution operator in the flavor basis is given by%
\begin{equation}
\overline{U}_{j}(x_{j},x_{j-1})=\left\{
\begin{array}{cc}
\cos \widetilde{\theta } & \sin \widetilde{\theta } \\
-\sin \widetilde{\theta } & \cos \widetilde{\theta }%
\end{array}%
\right\} \overline{U}_{j}^{P}\left\{
\begin{array}{cc}
\cos \widetilde{\theta } & -\sin \widetilde{\theta } \\
\sin \widetilde{\theta } & \cos \widetilde{\theta }%
\end{array}%
\right\}
\end{equation}%
where%
\begin{equation*}
\sin 2\widetilde{\theta }=\frac{\Delta \sin 2\theta }{\varpi }
\end{equation*}%
and $\widetilde{\theta }$ denotes the effective mixing angle in matter. With
the definitions $\widetilde{k}_{m}=\frac{\varpi }{2E}$ , $c_{j}=\cos [%
\widetilde{k}_{m}(x_{j}-x_{j-1})/2]$ , $s_{j}=$ $\sin [\widetilde{k}%
_{m}(x_{j}-x_{j-1})/2]$, \ $\overline{U}_{j}(x_{j},x_{j-1})$ becomes [10]%
\begin{equation}
\overline{U}_{j}(x_{j},x_{j-1})=\left\{
\begin{array}{cc}
c_{j}+is_{j}\cos 2\widetilde{\theta } & -is_{j}\sin 2\widetilde{\theta } \\
-is_{j}\sin 2\widetilde{\theta } & c_{j}-is_{j}\cos 2\widetilde{\theta }%
\end{array}%
\right\} .
\end{equation}%
The first order term demand a more hard work and we will explicit only the
final expression
\begin{equation}
\delta U(x_{j},x_{j-1})=-\frac{i}{2}\sin 2\widetilde{\theta }\left\{
\begin{array}{cc}
C_{j}\sin 2\widetilde{\theta } & C_{j}\cos 2\widetilde{\theta }-iS_{j} \\
C_{j}\cos 2\widetilde{\theta }+iS_{j} & -C_{j}\sin 2\widetilde{\theta }%
\end{array}%
\right\} .
\end{equation}%
where%
\begin{equation*}
S_{j}=\sqrt{2}G_{F}\int\limits_{x_{j-1}}^{x_{j}}dx\delta N_{j}(x)\sin
\widetilde{k}_{m}(x-\overline{\overline{x}})
\end{equation*}%
\begin{equation*}
C_{j}=\sqrt{2}G_{F}\int\limits_{x_{j-1}}^{x_{j}}dx\delta N_{j}(x)\cos
\widetilde{k}_{m}(x-\overline{\overline{x}})
\end{equation*}%
and $\overline{\overline{x}}=(x_{j}-x_{j-1})/2$.

The partial evolution operator when taken in first approach order is not
automatically unitary, being necessary to normalize it. That unitarity is
guaranteed by the redefinition%
\begin{equation}
U_{j}^{N}(x)=\frac{\overline{U}_{j}+\delta U_{j}(x)}{B^{2}}
\end{equation}%
where%
\begin{equation*}
\frac{1}{B}=\frac{\sqrt{4+[S_{j}^{2}+C_{j}^{2}]\sin ^{2}2\widetilde{\theta }}%
}{2}
\end{equation*}

\bigskip

\qquad By the utilization of the evolution operator as a product of ordered
infinitesimal operators (13) we calculate the transition probability as a
function of distance in zeroth-order approximation for the same case
presented in section II, a linearly varying matter density. The results are
presented in Fig.1. We calculate the same quantity by utilization of the
method used in ref. [9] , the result is presented in Fig. 1 also for
comparison. For neutrinos travelling in matter with an arbitrary density
profiles we must to take into account that not necessarily $\overline{N_{j}}%
(x)=$ $N_{e}(\overline{x})$. As an illustrative example we apply the method
for a sinusoidally varying matter density $N_{e}(x)=$ $N_{core}(1+\sin
\kappa x)$. With the approximation $\frac{\sin \kappa \Delta x}{\Delta x}$ $%
\simeq 1$ where $\Delta x$ is the length of the shells in the neutrino
trajectory the relation $\overline{N_{j}}(x)=$ $N_{e}(\overline{x})$ can be
used. In Fig. 2 the transition probability as a function of distance for
neutrinos at energy $500MeV$ and the physical parameters used in previous
sections is presented. The oscillation \ length is $\frac{2\pi }{\kappa }%
=4000km$. The same quantity calculated by using Runge-Kuta method is
presented also for comparison.

\section{The three neutrino case}

In the three neutrino case de evolution equation becomes%
\begin{equation}
i\frac{d}{dx}\nu _{f}(x)=\frac{1}{2E}\left\{ UM^{2}U^{-1}+\widehat{A}%
\right\} \nu _{f}(x)
\end{equation}%
where $M^{2}$ is the 3x3 diagonal matrix of mass
\begin{equation}
\nu _{f}(x)=\left( \nu _{e}(x),\nu _{\mu }(x),\nu _{\tau }(x)\right) ^{T}%
\text{ ; }U=e^{i\psi \Lambda _{7}}e^{i\phi \Lambda _{5}}e^{i\omega \Lambda
_{2}}
\end{equation}%
and $\widehat{A}$ is the matrix which the only different from zero element
is $A_{11}=A(x)$. Defining the quantities $\Delta =m_{2}^{2}-m_{1}^{2}$ ; $%
\Sigma =m_{2}^{2}+m_{1}^{2}$ ; $\Delta _{1}=\Sigma -2m_{3}^{2}$ and $\Lambda
=\Sigma -\Delta $ $c2\omega $, for a constant matter density, in terms of
the field $\Phi (x)=e^{-i\psi \Lambda _{7}}\nu _{f}(x)$ the evolution
equation becomes%
\begin{equation*}
i\frac{d}{dx}\Phi (x)=\widetilde{H}\text{ }\Phi (x).
\end{equation*}%
where%
\begin{equation*}
\widetilde{H}=\frac{1}{2E}\left\{ e^{i\phi \Lambda _{5}}e^{i\omega \Lambda
_{2}}M\text{ }e^{-i\omega \Lambda _{2}}e^{-i\phi \Lambda _{5}}+\widehat{A}%
\right\}
\end{equation*}%
or
\begin{equation*}
\widetilde{H}=\frac{\Sigma +m_{3}^{2}+A}{6E}+\delta H
\end{equation*}%
where $\delta H$ is the relevant term for transitions which elements are
given by $\frac{h_{ij}}{2E}$
\begin{equation*}
h_{11}=-\frac{(\Delta _{31}+\Delta _{32})c^{2}\phi }{2}+\frac{\Delta
_{31}+\Delta _{32}}{3}-\frac{\Delta c2\omega c^{2}\phi }{2}+\frac{2A}{3}
\end{equation*}%
\begin{equation*}
h_{22}=-\frac{\Delta _{31}+\Delta _{32}}{6}+\frac{\Delta \text{ }c2\omega }{2%
}-\frac{A}{3}\text{ };\text{ }h_{12}=\frac{\Delta \text{ }s2\omega \text{ }%
c\phi }{2}
\end{equation*}%
\begin{equation*}
h_{33}=-\frac{(\Delta _{31}+\Delta _{32})s^{2}\phi }{2}+\frac{\Delta
_{31}+\Delta _{32}}{3}-\frac{\Delta c2\omega s^{2}\phi }{2}-\frac{A}{3}
\end{equation*}%
\begin{equation*}
h_{23}=-\frac{\Delta \text{ }s2\omega \text{ }s\phi }{2}\text{ ; }h_{13}=%
\frac{(\Delta _{31}+\Delta _{32}+\Delta \text{ }c2\omega )s2\phi }{4}.
\end{equation*}%
The eigenvalues of the hamiltonian $\widetilde{H}$ are the effective
quantities $\frac{\widetilde{m}_{i}^{2}}{2E}$ in presence of matter and are
related to de roots $\lambda _{i}$ of the hamiltonian $\delta H$ by the
relation%
\begin{equation*}
\frac{\widetilde{m}_{i}^{2}}{2E}=\frac{\Sigma +m_{3}^{2}+A}{6E}+\lambda _{i}.
\end{equation*}%
With the help of the quantities $Q$ and $R$
\begin{equation*}
Q=-\left\{ \frac{\Delta ^{2}}{4}+\frac{\Delta _{1}^{2}}{12}+\frac{A^{2}}{3}-%
\frac{A\Delta c^{2}\phi c2\omega }{2}+\frac{A\Delta _{1}(c^{2}\phi
-2s^{2}\phi )}{6}\right\}
\end{equation*}%
\begin{equation*}
R=\frac{-1}{27}\left\{ \frac{\Delta _{1}^{3}}{4}-\frac{9\Delta _{1}\Delta
^{2}}{4}+\frac{3A\Delta _{1}^{2}(c^{2}\phi -2s^{2}\phi )}{4}-\frac{%
3A^{2}\Delta _{1}(c^{2}\phi -2s^{2}\phi )}{2}\right\}
\end{equation*}%
\begin{equation*}
\frac{-1}{27}\left\{ \frac{9A^{2}\Delta c^{2}\phi c2\omega }{2}-\frac{%
9A\Delta ^{2}(c^{2}\phi -2s^{2}\phi )}{4}+\frac{9A\Delta _{1}\Delta
c^{2}\phi c2\omega }{2}-2A^{3}\right\}
\end{equation*}%
\begin{equation*}
\cos \alpha =\frac{-R}{2\sqrt{\frac{-Q^{3}}{27}}}
\end{equation*}%
the roots $\lambda _{i}$ results%
\begin{equation*}
\lambda _{1}=\frac{-1}{E}\sqrt{\frac{-Q}{3}}\cos \frac{\alpha }{3}
\end{equation*}%
\begin{equation*}
\lambda _{2}=\frac{1}{2E}\sqrt{\frac{-Q}{3}}\cos \frac{\alpha }{3}-\frac{1}{%
2E}\sqrt{-Q}\sin \frac{\alpha }{3}
\end{equation*}%
\begin{equation*}
\lambda _{3}=\frac{1}{2E}\sqrt{\frac{-Q}{3}}\cos \frac{\alpha }{3}+\frac{1}{%
2E}\sqrt{-Q}\sin \frac{\alpha }{3}.
\end{equation*}

The\ elements of the symmetric evolution operator $T(x,_{0})$ for the field $%
\Phi (x)$ can be found in terms of these roots [5] and result

\begin{equation*}
T_{11}=\sum\limits_{j}\frac{\left[ (\lambda _{j}-h_{22})(\lambda
_{j}-h_{33})-h_{23}^{2}\right] }{d_{jkl}}e^{-i\lambda _{j}x/2E}
\end{equation*}%
\begin{equation*}
T_{22}=\sum\limits_{j}\frac{\left[ (\lambda _{j}-h_{11})(\lambda
_{j}-h_{33})-h_{13}^{2}\right] }{d_{jkl}}e^{-i\lambda _{j}x/2E}
\end{equation*}%
\begin{equation}
T_{33}=\sum\limits_{j}\frac{\left[ (\lambda _{j}-h_{11})(\lambda
_{j}-h_{22})-h_{12}^{2}\right] }{d_{jkl}}e^{-i\lambda _{j}x/2E}
\end{equation}%
\begin{equation*}
T_{12}=\sum\limits_{j}\frac{\left[ (\lambda _{j}-h_{33})h_{12}+h_{13}h_{23}%
\right] }{d_{jkl}}e^{-i\lambda _{j}x/2E}
\end{equation*}%
\begin{equation*}
T_{23}=\sum\limits_{j}\frac{\left[ (\lambda _{j}-h_{11})h_{23}+h_{12}h_{13}%
\right] }{d_{jkl}}e^{-i\lambda _{j}x/2E}
\end{equation*}%
\begin{equation*}
T_{13}=\sum\limits_{j}\frac{\left[ (\lambda _{j}-h_{22})h_{13}+h_{12}h_{23}%
\right] }{d_{jkl}}e^{-i\lambda _{j}x/2E}.
\end{equation*}%
where $d_{jkl}=(\lambda _{j}-\lambda _{k})(\lambda _{j}-\lambda _{l})$, with
$j\neq k\neq l$, and the evolution operator in the flavor basis equals:%
\begin{equation}
G(x,x_{0})=e^{i\psi \Lambda _{7}}T(x,x_{0})e^{-i\psi \Lambda _{7}}.
\end{equation}

For varying matter density the strategy is the same of the expression (13)
taking the evolution operator at zeroth order of approximation in the jth
shell given by expression (27). For more practical numerical calculations is
most easy to calculate the evolution of the field $\Phi (x)$ when the
partial operator is $T(x_{j},x_{j-1})$ and after to obtain the amplitudes $%
\nu _{f}(x)$ with the relation between the fields $\nu _{f}(x)$ and $\Phi (x)
$.In the figure [1] we present the transition probability $P_{ex}=P_{e\mu }$
$+$ $P_{e\tau }$ as a function of distance in zeroth-order approximation in
the three neutrino case (TNC) for a linearly varying matter density, the
mixing angle $\omega =28.7^{0}$ and the difference of squared mass $\Delta
_{21}=5\times 10^{-5}eV^{2}$ are chosen from indications of Solar neutrinos%
\textbf{\ }[19]. In the figure [3] the transition probability $P_{ex}$
computed by using the approximate evolution operators (13) and the partial
operators $T(x_{j},x_{j-1})$ given by (27) is presented for neutrinos at
energy $E=400$ $MeV$ and the same parameters of the figure [1], the
probabilities $P_{e\mu }$ and $P_{e\tau }$ are presented also. In the
figure[4], for neutrinos at energy $E=500MeV$ the probabilities $P_{ex}$
calculated by using expressions (13) and (27) and by using the exact
solution (27) are presented for neutrinos travelling in a constant matter
density $N_{e}(x)=1.5$ $N_{core}$ . The case $\omega =7^{0}$ is presented
also.

\section{Summary and conclusions}

We have presented approximate solutions to the neutrino evolution equation
calculated by different methods. The Fig. 1 shows the agreement between
numerical results and that one obtained by using the strategy of Elisio Lisi
and Daniele Montanino and a little disagreement between these results and
that one calculated using the strategy of Mattias Blennow and Tommy Ohlsson
for $L\approx 3000$ $km$. One can see in ref [9] that the convergence of the
transition probability calculated by using series solutions with numerical
results is better as much terms are included in these series and for
different numbers of terms the disagreement occur at different values of $L$%
. The results calculated by the utilization of the evolution operator as a
product of ordered partial operators at first-order approximation do not
introduce significative differences and are not presented. Additionally, for
an arbitrary density profile using the strategy of Lisi at zeroth-order
approximation is necessary only to calculate the appropriate $\overline{N_{j}%
}(x)$ according to we can see in the figure[2] for a sinusoidally varying
matter density. In contrast, by using the strategy of Blennow is necessary a
lot of calculations for each new law of variation density. The differences
of to calculate the approximate evolution operators as a product of ordered
partial operators in a two and three neutrino cases are clearly showed in
the figure [3] where $P_{ex}$ is presented for neutrinos at $E=400$ $MeV$.
In the three neutrino case the calculation of any quantity $P_{\alpha \beta }
$ is performed directly from expression (27). In the figure[4] we have
superposition of the probabilities calculated by using ordered partial
operators and the exact solution to the neutrino evolution equation and the
case of $\omega =7^{0}$ tend to the two neutrino case in order to
demonstrate the consistency of the method. Meanwhile the main question from
this figure is: For constant matter density $N_{e}(x)=1.5$ $N_{core}$ and
the parameters used in this case, the relation $\widetilde{\Delta }_{31}>>%
\widetilde{\Delta }_{21}$ is real but there is a disagreement between $P_{ex}
$ calculated by using the two and three neutrinos scenarios. As the neutrino
penetrate more and more in the matter, the initial little differences will
propagate along of the neutrino trajectory and the disagreement between
these approaches will appear for great values of $L$. For neutrinos at $%
E=1200$ $MeV$ travelling across the Earth matter this effect will appear
also in agreement of the results presented in the figure [1] near $L\simeq
3000Km$. Finally, we have derived an approximative expression to the
evolution operator for a system of three flavors of neutrinos propagating in
matter. Differently of the approaches used in references [12],[13],[14] and
[15] the method can be used for any values of the neutrino energy, mixing
angles in vacuum and difference of squared masses and applied to neutrinos
propagating at matter with an arbitrary density profile. The expressions to
the components of the evolution operator in a constant matter density (27),
from reference [5], are more compact and practical for to construct the
partial operators, than ones from the reference [6]. Due to the form of the
matrix $\widehat{A}$ of interaction between neutrinos and matter, the
extension for to introduce CP violation needs only a little modification in
the expression (27).

\textbf{References}

[1] Rabindra N. Mohapatra and P. B. Pal, Massive Neutrinos in Physics and
astrophysics, World Scientific Lectures Notes in Physics \textbf{41}, (1991)

[2] L. Wolfenstein, Phys. Rev. D \textbf{17}, 2369 (1978); 20, 2364 (1979).
S.P.Mikheyev and A. Yu. Srmirnov, Sov. J. Nucl. Phys. \textbf{42}, 1913
(1985); Nuovo Cimento \textbf{9C}, 17 (1986)

[3] Barger V. et al., Phys. Rev. D \textbf{22}, 2718 (1980)

[4] H. W. Zaglauer, Phys. Lett. B \textbf{198}, 556 (1987);

[5] V.M. Aquino et. al., Brazilian Journal of Physics, \textbf{27}, 384\
(1997)

[6]Tommy Ohlsson and Hakan Snellmann, hep-ph/9910546 (1999)

[7] P. B. Pal, Int. J. of Mod. Phys. A \textbf{7}, 5387 (1992)

[8] T. K. Kuo and James Pantaleone, Phys. Rev. D \textbf{35}, 3432 (1987)

[9] Mattias Blennow and Tommy Ohlsson, hep-ph/0405033 v2 26jul2004

[10] Elisio Lisi and Daniele Montanino, Physical Rev. D \textbf{56}, 1792
(1997)

[11] O.L.G. Peres and A. Yu. Smirnov, Physics Letters B \textbf{456}, 204
(1999)

[12] O.L.G. Peres and A. Yu. Smirnov, Nuclear Physics B \textbf{680}, 479
(2004)

[13] E. Kh. Akhmedov, M. Maltoni and A. Yu. Smirnov, Physical Review Letters
\textbf{95}, 211801-1 (2005)

[14] A. N. Ioannisian, N.A. Kazarian, A. Yu. Smirnov, and D. Wyler,
Pshysical Review D \textbf{71}, 033006-1 (2005)

[15] G.L. Fogli, G. Lettera and E. Lisi, hep-ph/0112241 (2001)

[16] Tommy Ohlsson and Hakan Snellmann, Physics Letters B \textbf{474}. 153
(2000)

[17] S. P. Mikheyev and A. Yu. Smirnov, Sov. J. Nucl. Phys. \textbf{42},
1913 (1985)

[18] I. Fujivara, Prog. Theor. Phys.\textbf{\ 7}, 433 (1952)

[19] SNO Collaboration, Q.R. Ahmad, et al., nucl-ex/0309004.

Acknowledgements

This work was supported by CPG-UEL and CNPq

Figure Caption 1

The neutrino oscillation probability $P_{ex}$ for a linearly varying matter
density as a function of the distance for $E=1.2GeV$. The solid black line
corresponds to the numerical solution, the red line to the series solution
and the blue line and the black dashed line are the results obtained using
the evolution operator as a product of ordered partial operators at
zeroth-order approximation in the two and three neutrino cases respectively.

Figure Caption 2

The neutrino oscillation probability $P_{ex}$ for a sinusoidally varying
matter density $N_{e}(x)=$ $N_{core}(1+\sin \kappa x)$ as a function of the
distance. The neutrino energy is $500MeV$ and the oscillation \ length is $%
\frac{2\pi }{\kappa }=4000km$. The solid red line corresponds to the
numerical solution and the blue line is the results obtained using the
evolution operator as a product of ordered partial operators at zeroth-order
approximation.

Figure Caption 3

The neutrino oscillation probability $P_{ex}$ for a linearly varying matter
density as a function of the distance for $E=400GeV$. The black line and the
red line correspond to the results obtained using the evolution operator as
a product of ordered partial operators at zeroth-order approximation in the
two and three neutrino cases respectively. The blue \ and green lines are
the probabilities $P_{e\mu }$ and $P_{e\tau }$ obtained in a three neutrino
context with the angle $\psi =0$.

Figure Caption 4

The neutrino oscillation probability $P_{ex}$ for a constant matter density
as a function of the distance for $E=500$ $MeV$. The black line corresponds
to the results obtained using the evolution operator as a product of ordered
partial operators at zeroth-order approximation in the two neutrino case.
The red line and the blue dashed line are the probabilities $P_{ex}$
obtained in a three neutrino context approximately and exactly. The green
line corresponds to the approximative case with $\omega =7^{0}$.

\end{document}